\documentclass[twocolumn,amsmath,amssymb]{revtex4}
\usepackage{graphicx}
\usepackage{dcolumn}
\usepackage{color}
\usepackage{bm}
\usepackage{amsmath,amssymb,graphicx}
\usepackage{epsfig}
\usepackage{enumerate}
\usepackage{array}
\usepackage{multirow}

\begin{document}
\title
{Generalised energy equipartition in electrical circuits}
\author{Aritra Ghosh\footnote{ag34@iitbbs.ac.in}}
\affiliation{School of Basic Sciences,\\ Indian Institute of Technology Bhubaneswar, Argul, Jatni, Khurda, Odisha 752050, India}
\vskip-2.8cm
\date{\today}
\vskip-0.9cm


\begin{abstract}
In this brief note, we demonstrate a generalised energy equipartition theorem for a generic electrical circuit with Johnson-Nyquist (thermal) noise. From quantum mechanical considerations, the thermal modes have an energy distribution dictated by Planck's law. For a resistive circuit with some inductance, it is shown that the real part of the admittance is proportional to a probability distribution function which modulates the contributions to the system's mean energy from various frequencies of the Fourier spectrum. Further, we analyse the case with a capacitor connected in series with an inductor and a resistor. The results resemble superstatistics, i.e. a superposition of two statistics and can be reformulated in the energy representation. The correct classical limit is obtained as $\hbar \rightarrow 0$.
\end{abstract}

\maketitle

\section{Introduction}
 In the context of electrical engineering, noise can be understood to be an unwanted disturbance in an electrical signal. Perhaps the earliest systematic study of noise in electrical circuits is due to Schottky \cite{shotnoise}, who described what is now known as `shot' noise. In the following decade, Johnson \cite{johnson1,johnson2}, and subsequently Nyquist \cite{nyquist}, described `thermal' noise in electrical circuits, which originates from the random thermal motion of charge carriers (typically electrons) in a conductor at any finite temperature. The Johnson-Nyquist noise is therefore, unavoidable at ordinary temperatures, except for when the conductor is cooled to cryogenic temperatures. In the latter case, although thermal noise may become negligible, quantum noises (such as shot noise) arising from zero-point fluctuations play an important role. It should be remarked that the notion of Johnson-Nyquist noise is of practical interest in thermometry \cite{thermo}.\\

Mathematically, a noise \(V(t)\) is a stationary random process which can only be characterised by its statistical properties. Obviously, due to its randomness, one has \(\langle V(t) \rangle = 0\), where the mean \(\langle \cdot \rangle\) is taken over all noise realisations. The autocorrelation function \(C_V(\tau) = \langle V(t+\tau) V(t)\rangle\) characterises the statistical properties of the noise. Note that \(C_V(\tau)\) only depends on the time difference \(\tau\) and not \(t\). This is due to the fact that the process is stationary and is therefore time-homogenous. If \(C_V(\tau) \sim \delta(\tau)\), then the noise at any instant of time \(t\) is uncorrelated with the noise at any other time instant \(t' \neq t\). Such a noise is known as a white noise because the spectral density is a constant, independent of frequency. However, in practical situations, noises are rarely white, and a general form of the intensity spectrum is \(S_V(\omega) \sim 1/\omega^\alpha\). For \(\alpha = 0\), one gets the usual white noise, whereas the cases with \(\alpha = 1\) and \(\alpha =2\) are termed as pink and red (or, Brown) noises respectively \cite{noise1,noise2}. \\

The spectral density or power spectrum of thermal noise taking into account the quantum mechanical nature of electrons was given by Nyquist \cite{nyquist} and has the following expression:
\begin{equation}
S_V (\nu) d\nu = \frac{4 R_\nu h \nu d\nu }{e^{h\nu/k_B T} - 1},
\end{equation} where \(R_\nu\) is the frequency dependent resistance of the circuit. Using the definition \(\omega = 2 \pi \nu\), the power spectrum takes the following form:
 \begin{equation}\label{Snoise}
S_V (\omega) d\omega  = \frac{(2/\pi) R(\omega) \hbar \omega d\omega }{e^{\hbar \omega /k_B T} - 1}.
\end{equation} It is a one-dimensional analogue of Planck's law \cite{bb}. Let us note that due to the non-trivial frequency dependence of \(S_V(\omega)\), thermal noise is not white. However, for a frequency independent \(R(\omega)\) (which is often approximately true) and \(k_B T >> \hbar \omega\), one has
\begin{equation}
S_V (\omega) \approxeq \frac{2 R k_B T}{\pi} ,
\end{equation} and in that case, the spectral density is a constant and the thermal noise is approximately white. This is often the case in a typical experimental setting \cite{johnson1,johnson2}. \\

There is an intriguing analogy between the thermal motion of electrons in a conductor and the motion of a Brownian particle immersed in a fluid \cite{brown1,brown3,brown5,brown7}. Both electron and Brownian particle possess some kinetic energy due to their temperature, suffer frequent collisions and their trajectories are zig-zag. Thus, a mathematical model mimicking that of Brownian motion is expected to describe the phenomena of thermal noise, at least at a phenomenological level \cite{brown7}. To this end, let us consider a closed loop with some resistance \(R\) and an overall inductance of the loop given by \(L\). Then, the thermal noise \(V(t)\) at any temperature \(T\) would drive a small time-varying current through the loop. From the Kirchhoff's voltage equation, we may write
\begin{equation}\label{eom}
L \frac{dI(t)}{dt} + RI(t) = V(t),
\end{equation} where \(I(t)\) is the current. Since \(V(t)\) is a noise, the above voltage equation resembles the equation of motion of a Brownian particle, where \(V(t)\) is replaced by the fluctuating force \(\eta(t)\) on the Brownian particle due to random bombardment by molecules of the surrounding medium. In this paper, we shall formulate a generalisation of classical energy equipartition theorem in the context of an electrical circuit with thermal noise. These results would also be analysed from the point of view of superstatistics.\\

The rest of the paper is organised as follows. In the next section [section-(\ref{presec})], we briefly discuss some mathematical notions which will aid our analysis. Following this, in section-(\ref{main}), we compute the mean energy in a generic circuit with thermal noise. Our main result on a generalised energy equipartition theorem is presented. The superstatistics viewpoint is discussed in section-(\ref{ssec}). Finally, we conclude the paper with some discussion in section-(\ref{discuss}). 

\section{Preliminaries}\label{presec}
Let us first describe some necessary mathematical tools for our analysis. However, we shall be quite brief and the reader is referred to \cite{brown7} and references therein, for more details. Consider a stationary stochastic process \(X = X(t)\) such that \(\langle X(t) \rangle = 0 \). By stationary, we mean that all the equal-time moments of \(X(t)\) are independent of \(t\). The autocorrelation function is defined as
\begin{equation}\label{corr}
C_X (\tau) = \langle X(t + \tau) X(t) \rangle,
\end{equation} which admits a Fourier decomposition:
\begin{equation}\label{corr1}
C_X (\tau) = \int_0^\infty S_X(\omega) \cos (\omega \tau) d\omega, \hspace{5mm} \tau \geq 0.
\end{equation} Here, \(S_X (\omega)\) is the positive-frequency Fourier amplitude, often called the spectral density or intensity spectrum, which is given by
\begin{equation}\label{ftc}
S_X (\omega) = \frac{2}{\pi} \int_0^\infty C_X(\tau) \cos (\omega \tau) d\tau, \hspace{5mm} \omega \geq 0.
\end{equation}
The fact that the autocorrelation function and the spectral density are Fourier (cosine) transforms of each other is known as the Wiener-Khintchine theorem \cite{brown7,WKref1,WKref2}. It may be easily verified that for \(C_V (\tau) \sim \delta (\tau)\), one finds \(S_V(\omega)\) to be a constant, independent of frequency. This is the case of white noise. Moreover, if we put \(\tau = 0\) in eqns (\ref{corr}) and (\ref{corr1}), we get
\begin{equation}\label{cx}
\langle X (t)^2 \rangle = \int_0^\infty S_X(\omega) d\omega. 
\end{equation}
With this background, we may begin our main analysis. 

\section{Mean energy in a circuit with thermal noise}\label{main}
\subsection{\(LR-\)circuit}
Consider a generic electrical circuit with an inductance \(L\) and some resistance \(R\). At a finite temperature, there is a thermal voltage or noise \(V(t)\) which drives a current \(I(t)\) in the loop. The equation of motion for the loop is given by eqn (\ref{eom}). We may solve this differential equation using Fourier-Laplace transforms. Then eqn (\ref{eom}) gives the algebraic equation:
\begin{equation}\label{VI}
\tilde{I} (\omega) = \mathcal{A}(\omega) \tilde{V}(\omega) ,
\end{equation} where, `tilde' denotes Fourier-Laplace transform, and
\begin{equation}\label{admit}
\mathcal{A}(\omega) = \frac{1}{- i \omega L + R}
\end{equation} is the admittance. It follows from eqn (\ref{VI}) that
\begin{equation}
S_I (\omega) = |\mathcal{A}(\omega)|^2 S_V(\omega) .
\end{equation} Thus, from eqn (\ref{corr1}), the current autocorrelation function reads:
\begin{equation}
C_I (\tau) = \langle I(t + \tau) I(t) \rangle = \int_{0}^\infty  |\mathcal{A}(\omega)|^2 S_V (\omega) \cos (\omega \tau) d\omega.
\end{equation}
We may set \(\tau =0\) without loss of generality. Then from eqn (\ref{Snoise}), one has
\begin{eqnarray}
 \langle I(t)^2 \rangle &=& \frac{2}{\pi} \int_0^\infty  \frac{R |\mathcal{A}(\omega)|^2 \hbar \omega }{e^{\hbar \omega /k_B T} - 1} d\omega \nonumber \\
 &=& \frac{2}{\pi} \int_{0}^\infty \frac{R}{L^2 \omega^2 + R^2}  \frac{ \hbar \omega }{e^{\hbar \omega /k_B T} - 1} d\omega. 
\end{eqnarray}
The above equation is a fluctuation-dissipation theorem \cite{kubo,callen}, relating the autocorrelation function of a dynamical variable to the response function (here, the admittance). Now, the energy is
\begin{equation}\label{E}
E := \frac{L \langle I^2 \rangle}{2} = \frac{1}{\pi} \int_{0}^\infty \frac{R/L}{\omega^2 + (R/L)^2}  \frac{ \hbar \omega }{e^{\hbar \omega /k_B T} - 1} d\omega,
\end{equation} where we have suppressed \(t\) from the left-hand side because the result does not depend on \(t\) owing to stationarity of the process. Eqn (\ref{E}) is a quantum generalisation of equation (1) of Johnson's original paper \cite{johnson2}. An interesting interpretation can be ascribed to the above result. If we put
 \begin{equation}\label{epsilon}
 \epsilon(\omega,T) = \frac{\hbar \omega/2}{e^{\hbar \omega /k_B T} - 1},
 \end{equation} and identify, 
\begin{equation}
P_0(\omega) := \frac{2L{\rm Re} [\mathcal{A}(\omega)]}{\pi} =  \frac{2}{\pi} \frac{R/L}{\omega^2 + (R/L)^2} ,
\end{equation} then, eqn (\ref{E}) takes the following form:
\begin{equation}\label{Eavg}
E := \frac{L \langle I^2 \rangle}{2} = \int_{0}^\infty P_0(\omega) \epsilon (\omega,T) d\omega.
\end{equation} Therefore, it seems that the mean energy in the circuit is an average over the distribution function \(P_0(\omega)\), which is clearly positive definite (because \(R > 0\)). It is this function \(P_0(\omega)\) which is proportional to the real part of the admittance that controls which frequencies contribute more to the mean energy as compared to others. It simply follows that
\begin{eqnarray}
\int_{0}^\infty P_0(\omega) d\omega &=& \frac{2}{\pi} \int_{0}^\infty \frac{(R/L) d\omega}{\omega^2 + (R/L)^2} \nonumber \\
 &=& \frac{1}{\pi} \int_{-\infty}^\infty \frac{(R/L) d\omega}{\omega^2 + (R/L)^2} \nonumber \\
 &=& 1,
\end{eqnarray} or equivalently, the function \(P_0(\omega)\) is a probability distribution function in the Fourier space. Therefore, one may interpret \(P_0(\omega) \epsilon (\omega,T) d\omega\) as the energy contribution to the mean energy of the circuit coming from the frequency interval from \(\omega\) to \(\omega + d\omega\). \\

Let us note that the mean energy of the circuit is expressible as a two-fold average. The first averaging is over the thermal state of the charge carriers leading to \(\epsilon(\omega,T)\). The second averaging is explicit in eqn (\ref{Eavg}) wherein the averaging takes place over the Fourier spectrum. The latter is modulated by a suitable probability distribution function \(P_0(\omega)\) which is proportional to the real part of the admittance. The careful reader should have noted that \(\epsilon (\omega,T)\) is actually only 1/2 the energy of a quantum oscillator. This is precisely because an oscillator has both potential and kinetic energies (that are equal at thermal equilibrium) which means \(\epsilon(\omega,T)\) is just the mean energy per degree of freedom. \\

In the classical limit, one has \(\hbar \rightarrow 0\) and subsequently, \(\epsilon \rightarrow k_B T/2\). This means, eqn (\ref{Eavg}) gives
\begin{equation}
E := \frac{L \langle  I^2 \rangle}{2} = \frac{k_B T}{2} \int_{0}^\infty P_0(\omega) d\omega = \frac{k_B T}{2},
\end{equation} consistently \cite{brown7,waterson}. Thus, in a sense, eqn (\ref{Eavg}) generalises the energy equipartition theorem for an electrical circuit, taking into account quantum mechanical considerations. Eqn (\ref{Eavg}) is the electrical analogue of the recently proposed quantum counterpart of energy equipartition theorem wherein, the mean energy of a quantum Brownian particle is expressible as a similar two-fold average \cite{jarzy1,jarzy2,jarzy3,jarzy4,superstat,jarzy5,jarzy6,kaur,kaur2}. \\

\subsection{\(LCR-\)circuit}
Consider now a situation similar to that considered in the previous subsection but now with a capacitor with capacitance \(C\) in series with a resistance \(R\). The inductance of the loop is \(L\). Thermal noise \(V(t)\) drives an electric current in the circuit. Using the definition \(I(t) = dQ(t)/dt\), we have the following differential equation for the loop from Kirchhoff's voltage rule:
\begin{equation}
L \frac{d^2 Q(t)}{dt^2} + R \frac{dQ(t)}{dt} + \frac{Q(t)}{C} = V(t),
\end{equation} where \(\langle V(t) \rangle = 0\). As before, we shall take a Fourier-Laplace transform of the above equation so that we get
\begin{equation}
\tilde{Q}(\omega) = \mathcal{X} (\omega) \tilde{V}(\omega),
\end{equation} where, 
\begin{equation}
\mathcal{X} (\omega)= \frac{1}{- \omega^2 L + (1/C) - i \omega R},
\end{equation} is a suitable response function. The spectral densities \(S_Q (\omega)\) and \(S_V(\omega)\) are related as
\begin{equation}
S_Q (\omega) = |\mathcal{X} (\omega) |^2 S_V (\omega) ,
\end{equation} which means
\begin{eqnarray}
\langle Q(t + \tau) Q(t) \rangle &=& \int_0^\infty |\mathcal{X} (\omega) |^2 S_V (\omega) \cos (\omega \tau) d\omega \nonumber \\\
&=& \int_0^\infty \frac{(1/L^2)S_V (\omega) \cos (\omega \tau) d\omega}{ (\omega_0^2 - \omega^2)^2 + \omega^2 (R/L)^2},  \label{1234}
\end{eqnarray} where we have defined \(\omega_0 = 1/\sqrt{LC}\). \\

The electrical energy stored in the capacitor is defined as \(E_C = \langle Q^2 \rangle/2 C\). At this stage, we put \(\tau = 0\) to obtain
\begin{eqnarray}
\langle Q^2 \rangle &=& \int_0^\infty \frac{(1/L^2)S_V (\omega) d\omega}{ (\omega_0^2 - \omega^2)^2 + \omega^2 (R/L)^2}  \nonumber \\\
&=& \frac{2R}{\pi L^2}  \int_0^\infty \frac{1}{ (\omega_0^2 - \omega^2)^2 + \omega^2 (R/L)^2}  \frac{\hbar \omega d\omega}{e^{\hbar \omega /k_B T} - 1}. \nonumber \\
\end{eqnarray} 
Thus, the mean energy of the capacitor is given by
\begin{equation}\label{ECAvg}
E_C :=  \frac{\langle Q^2 \rangle}{2 C} = \int_0^\infty P_C(\omega) \epsilon(\omega,T) d\omega ,
\end{equation} where, \(\epsilon(\omega,T)\) is the mean energy per degree of freedom of a thermal mode with frequency \(\omega\) and temperature \(T\) [eqn (\ref{epsilon})], while
\begin{equation}
 P_C(\omega) =   \frac{(2R \omega_0^2/\pi L)}{ (\omega^2 - \omega_0^2)^2 + \omega^2 (R/L)^2} .
\end{equation}
Obviously, \(P_C(\omega)\) is positive definite which is clear by inspection. It is easy to verify by explicit integration that
\begin{equation}
\int_0^\infty P_C(\omega) d\omega = \frac{R \omega_0^2}{\pi L} \int_{-\infty}^\infty \frac{d\omega}{(\omega^2 - \omega_0^2)^2 + \omega^2 (R/L)^2} = 1.
\end{equation}
Thus, the function \(P_C(\omega)\) can be interpreted as a probability distribution function which modulates the energy stored in the capacitor in the sense that \(  P_C (\omega) \epsilon(\omega,T) d\omega\) is the portion of the mean energy of the capacitor from the frequency interval from \(\omega\) to \(\omega + d\omega\). \\

Next, let us find an expression for the mean energy in the inductive element, i.e. \(E_L = L \langle I^2 \rangle/2\). We can obtain the correlation function \(\langle I ( t + \tau) I (t) \rangle \) by differentiating eqn (\ref{1234}) twice, to obtain
\begin{eqnarray}
\langle I(t + \tau) I(t) \rangle &=& \int_0^\infty \frac{(\omega^2/L^2) S_V (\omega) \cos (\omega \tau) d\omega}{ (\omega_0^2 - \omega^2)^2 + \omega^2 (R/L)^2}.  \label{12345}
\end{eqnarray}
Putting \(\tau=0\) and substituting eqn (\ref{Snoise}), we get
\begin{equation}\label{ELAvg}
E_L :=  \frac{L \langle I^2 \rangle}{2} = \int_0^\infty P_L(\omega) \epsilon(\omega,T) d\omega ,
\end{equation} where, 
\begin{equation}
P_L(\omega) = \frac{(2R \omega^2/\pi L)}{ (\omega^2 - \omega_0^2)^2 + \omega^2 (R/L)^2} .
\end{equation} The positivity of \(P_L(\omega)\) is once again clear by inspection and its normalisation is verified by an explicit integration, giving
\begin{equation}
\int_0^\infty P_L(\omega) d\omega = \frac{R}{\pi L} \int_{-\infty}^\infty \frac{\omega^2 d\omega}{(\omega^2 - \omega_0^2)^2 + \omega^2 (R/L)^2} = 1.
\end{equation} Thus, in the same sense as before, \(P_L(\omega)\) is a probability distribution function whose significance is that \( \epsilon(\omega,T) P_L (\omega) d\omega\) is the portion of the mean energy of the inductor from the frequency interval from \(\omega\) to \(\omega + d\omega\). Therefore, even in the \(LCR-\)circuit, the generalised energy equipartition theorem holds good, individually for the capacitor and the inductor. However, since \(P_C(\omega)\) and \(P_L(\omega)\) are different functions, the mean energies of the capacitor and the inductor are distributed in different ways across the Fourier spectrum. The total energy of the circuit is just a linear sum of the two. The distribution functions \(P_C(\omega)\) (red) and \(P_L(\omega)\) (blue) have been plotted in figure-(1) in dimensionless form and it is clearly seen that the the mean energies \(E_C\) and \(E_L\) receive non-uniform contributions from the thermal modes of different frequencies. In fact, \(E_C\) is found to be dominated by contributions from low frequency modes, i.e. \(\omega << R/L\) while \(E_L\) receives dominant contributions from thermal modes of frequency \(\omega \approx \omega_0 = R/2L\). For the sake of comparison, we have also plotted \(P_0(\omega)\) (black-dashed line) describing the mean energy of the \(LR-\)circuit in figure-(1). It can be quickly verified that as \(\hbar \rightarrow 0\), one obtains \(E_C = E_L = k_BT/2\) consistent with our expectations. \\

\begin{figure}
	\centering
		\includegraphics[width=3in]{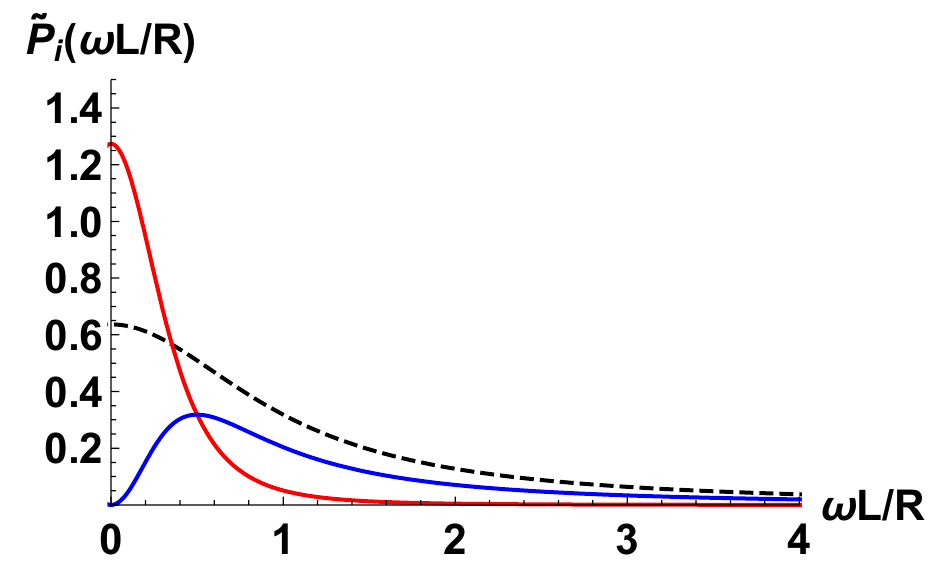}			
		
		\caption{\footnotesize Plot of dimensionless distribution functions \(\tilde{P}_i(\omega L/R) = (R/L) P_i(\omega L/R)\) as a function of dimensionless frequency \(\omega L/ R\). Here the index \(i = 0\) (black-dashed), \(i = C\) (red-solid) and \(i=L\) (blue-solid), respectively implies the functions \(P_0(\omega)\), \(P_C(\omega)\) and \(P_L(\omega)\). For the latter two, we have put \(\omega_0 = R/2L\) or equivalently, \(C = L/\pi^2 R^2\).}
	
\end{figure}

\section{Superstatistics of energy}\label{ssec}
We have already computed the mean energies of inductive and capacitive circuit elements in the previous section and a generalised energy equipartition theorem was proposed in this context. One should note that the mean energies are obtained as a two-fold average, reminiscent of superstatistics, i.e. superposition of two statistics \cite{ss1,superstat}. For instance, in case of the \(LR-\)circuit, the mean energy [eqn (\ref{Eavg})] is obtained as: (i) averaging over the thermal state of the environment to obtain \(\epsilon(\omega,T)\), (ii) averaging over all the frequencies using distribution function \(P_0(\omega)\).  \\

Following the analysis performed in \cite{superstat} on the energetics of a free quantum Brownian particle, we may re-express eqn (\ref{Eavg}) (and similarly eqns (\ref{ECAvg}) and (\ref{ELAvg})) in the energy representation. Consider eqn (\ref{Eavg}): 
\begin{equation}
E = \int_0^\infty \epsilon(\omega,T) P_0 (\omega) d\omega = \int_{k_B T/2}^\infty \epsilon P_0 (\omega(\epsilon,T)) \frac{d\omega}{d\epsilon} d\epsilon,
\end{equation} where in the second equality above, we have inverted \(\epsilon(\omega,T)\) to give \(\omega = \omega(\epsilon,T)\). If we now define 
\begin{equation}
f (\epsilon,T) := P_0 (\omega(\epsilon,T)) \frac{d\omega}{d\epsilon},
\end{equation} then, 
\begin{equation}
E = \int_{k_B T/2}^\infty \epsilon f(\epsilon,T)  d\epsilon.
\end{equation} Thus, we have expressed the mean energy of the circuit in the energy representation using a new distribution function \(f(\epsilon,T)\). Since \(\int_{k_BT/2}^\infty f(\epsilon,T) d\epsilon = \int_0^\infty P_0 (\omega) d\omega = 1\), the new distribution function \(f(\epsilon,T)\) is normalised and the mean energy of the circuit is expressible in the form of an average over the thermal mode energies \(\epsilon\). One should note that the function \(f(\epsilon,T)\) depends on the temperature, unlike \(P_0(\omega)\). This distribution function \(f(\epsilon,T)\) has been estimated numerically and has been plotted as a function of \(\epsilon\) in figure-(2). It should be remarked that for the sake of convenience in the numerical analysis, we have included the ground state (kinetic) energy \(\hbar \omega/4\) in eqn (\ref{epsilon}) describing \(\epsilon(\omega, T)\). Now from figure-(2), it is clear that the distribution function falls off for large \(\epsilon\) meaning that the higher thermal mode energies contribute less to the mean energy of the circuit. An interesting behaviour is noted as \(\epsilon \rightarrow k_BT/2\), which is the lower bound on the values \(\epsilon\) can take. The function \(f(\epsilon,T) \rightarrow \infty\) as \(\epsilon \rightarrow k_BT/2\). This is the classical contribution to the mean energy of the circuit, for which \(d\epsilon/d\omega = 0\) leading to this divergence. The subsequent contributions are of purely quantum mechanical origin and ensure that the mean energy of the circuit taking into account quantum mechanical considerations always exceeds that of the classical counterpart, i.e. \(E \geq k_BT/2\) where the equality holds for \(\hbar \rightarrow 0\). A similar analysis is possible for the \(LCR-\)circuit but here we do not pursue it further. \\

\begin{figure}
	\centering
		\includegraphics[width=3in]{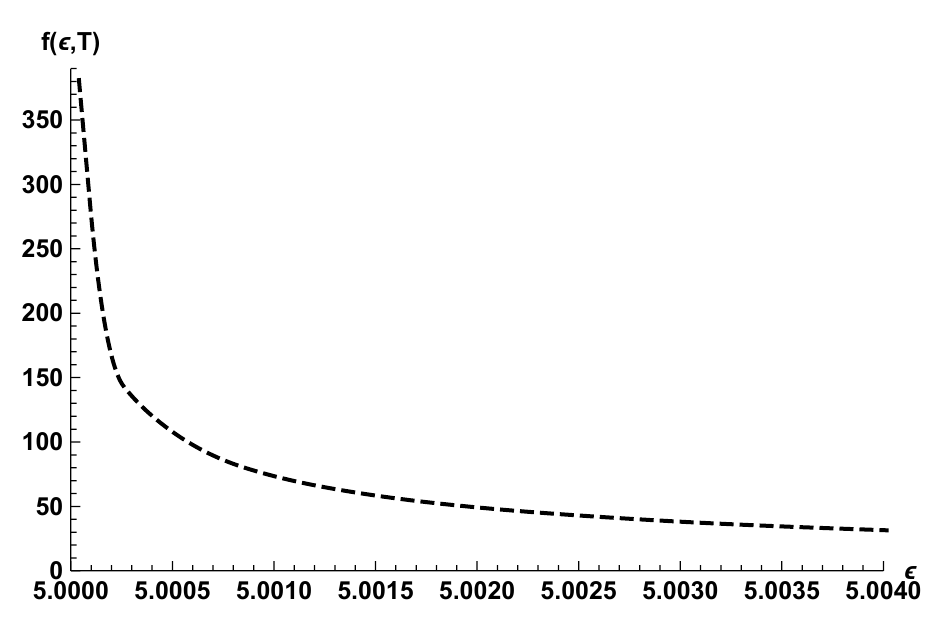}			
		
		\caption{\footnotesize Plot of the distribution function \(f(\epsilon,T)\) as a function of \(\epsilon\) for \(T = 10\). We have set \(\hbar = k_B = 1\) in the numerical analysis.}
	
\end{figure}

\section{Discussion}\label{discuss}
We have described a generalised quantum counterpart of energy equipartition theorem for an electrical circuit with thermal noise and have shown that as \(\hbar \rightarrow 0\), the correct classical result emerges. Central to our result is the existence of normalised probability distribution function(s) in the frequency domain such that the mean energy can be expressed as in eqn (\ref{Eavg}). These probability distribution functions \(\{P_i(\omega)\}\) control the contributions to the mean energy of the circuit (in the \(i\)th element) from various intervals in the Fourier spectrum in the sense that \(P_i(\omega) \epsilon(\omega,T)  d\omega\) is the part of the mean energy coming from the frequency interval \(\omega\) to \(\omega + d\omega\). For the \(LR-\)circuit, the function \(P_0(\omega)\) turns out to be proportional to the real part of the electrical admittance \(\mathcal{A}(\omega)\) which can be controlled in an experimental setting, and therefore \(P_0(\omega)\) is externally controllable. Further, it was emphasised that the mean energy obtained in this setting resembles superstatistics, i.e. a superposition of two statistics, and our results can be reformulated in the energy representation as demonstrated in section-(\ref{ssec}). \\

 In \cite{johnson1,johnson2,nyquist}, \(\mathcal{A}(\omega)\) corresponds to the transfer admittance between the element in which the thermal noise is generated and element in which it is measured. For our case, we do not describe measurements and assume the simple situation of a single closed loop with an inductance \(L\) and resistance \(R\), such that the thermal noise originates from random motion of electrons in the loop. Then \(\mathcal{A}(\omega)\) corresponds to the admittance of the loop as a whole. A Brownian motion-like analysis reveals that the total energy of the loop is distributed throughout the Fourier spectrum according to the probability distribution function \(P_0(\omega)\). This result was generalised with a capacitor in series and the same physical conclusions were obtained. It appears that the generalised energy equipartition theorem is a natural consequence of the Callen-Welton fluctuation-dissipation theorem \cite{kubo,callen} relating correlation functions of observables to relevant response functions. \\

To reiterate, thermal noise is due to the random motion of electrons in the circuit, and the thermal modes have been taken to be distributed according to Planck's law \cite{nyquist,bb}. It should be greatly emphasised that the present analysis relies on the Langevin equation and the existence of linear response functions (such as admittance). Henceforth, the present results are not expected to hold for circuits with non-linear and active elements such as diodes and transistors. 

\section*{Acknowledgements}
The author would like to thank Jasleen Kaur for several discussions, comments on the manuscript and for help in preparing the plots. The author also acknowledges Malay Bandyopadhyay for several related discussions. This work is supported by Ministry of Education (MoE), Government of India in the form of a Prime Minister's Research Fellowship (ID: 1200454).

\end{document}